\documentclass[twocolumn,showpacs,preprintnumbers]{revtex4}
\usepackage{amssymb}
\usepackage{amsmath}
\usepackage[dvips]{graphicx}
\usepackage[english]{babel}

\begin{document}

\title{Thermal fluctuations of vortex clusters in quasi two-dimensional
Bose-Einstein condensate}
\author{W. V. Pogosov and K. Machida}
\affiliation{Department of Physics, Okayama University, Okayama 700-8530, Japan}
\date{\today }

\begin{abstract}
We study the thermal fluctuations of vortex positions in small vortex
clusters in a harmonically trapped rotating Bose-Einstein condensate. It is
shown that the order-disorder transition of two-shells clusters occurs via
the decoupling of shells with respect to each other. The corresponding
"melting" temperature depends stronly on the commensurability between
numbers of vortices in shells. We show that "melting" can be achieved at
experimentally attainable parameters and very low temperatures. Also studied
is the effect of thermal fluctuations on vortices in an anisotropic trap
with small quadrupole deformation. We show that thermal fluctuations lead to
the decoupling of a vortex cluster from the pinning potential produced by
this deformation. The decoupling temperatures are estimated and strong
commensurability effects are revealed.
\end{abstract}

\maketitle

\section{Introduction}

Properties of Bose-Einstein condensates (BEC) of alkali atom gases attract a
considerable current interest. Recent progress in this field allows for the
creating of quasi-two-dimensional atomic gas either using $1D$ optical
lattices or applying tight axial trapping \cite%
{Ketterle,Schweikhard,Rychtarik,Smith}. It is well-known that, according to
the Mermin-Wagner-Hohenberg theorem, Bose-Einstein condensation is
impossible in $2D$ homogeneous systems in the thermodynamic limit. However,
a Bose-Einstein condensation at finite temperature becomes possible in a
trapped gas.

Recently, the Berezinskii-Kosterlitz-Thouless (BKT) transition associated
with the creation of vortex-antivortex pairs was studied theoretically in $%
2D $ BEC clouds \cite{Trombettoni,Simula,Baym} and it was shown that this
transition can occur in the experimentally attainable range of parameters.
For instance, according to Ref. \cite{Simula}, BKT transition can happen at $%
T\approx 0.5T_{c}$ for the number of particles $N\sim 10^{3}\div 10^{4}$ and
realistic values of other parameters. These results demonstrate an
importance of temperature effects in $2D$ BEC even at temperatures well
below the critical one. At the same time, the effect of temperature on
vortex lattices in BEC has not been studied yet, although fluctuation of
positions of vortices should become considerable even at lower temperatures
than those corresponding to the BKT transition. Finally, experimental
evidence for the BKT transition in trapped condensates was reported in Ref. 
\cite{Dalibard}. Recently, the effect of temperature on the vortex matter
was analyzed in Ref. \cite{Cast}, but in the strongly fluctuative regime at
relatively high temperatures, when positions of vortices are random.

It is well-known from the theory of superconductivity that thermal
fluctuations can lead to the melting of flux line lattice. However, in real
superconductors this usually happens only in the vicinity of the critical
temperature. For the case of atomic BEC, critical temperature depends on the
number of particles in the trap. Therefore, melting can occur at
temperatures much lower than the critical one. In finite systems,
fluctuations of vortex positions depend also on the number of vortices. In
such systems, melting temperature is not a strictly defined quantity. In
this case, a characteristic temperature of the order-disorder transition
("melting") can be defined using the Lindemann criterion, see the discussion
in \cite{Bedanov}. With increasing of the number of vortices, the
fluctuations of the vortex positions are determined by elastic shear modulus
of the system, i. e. by the Tkachenko modes studied in Ref. \cite{Tkachenko}%
. However, when a vortex number is not large, quantization effects start to
play a very important role and "melting" temperatures in this case can be
much smaller than that for a larger system. Thermal fluctuations of the
system of interacting point particles trapped by external potential were
studied before in Refs. \cite{Bedanov,Lozovik,Filinov,BedaPeeters,Peeters}
mostly by using Monte-Carlo simulations. If there are not many particles (or
vortices) in the system, in the ground state, they form a cluster consisting
of shells. It was shown in Refs. \cite{Lozovik,Filinov,BedaPeeters,Peeters}
that with increasing the temperature, first, an order between different
shells is destroyed and these shells become decoupled with respect to each
other. Only after this, with the sufficient increase of temperature, a
radial disordering of the cluster occurs. This leads to the hierarchy of
"melting" temperatures, which depends dramatically on the symmetry of
cluster and number of particles.

In addition to thermal fluctuations, quantum ones can be significant in
atomic condensate. In recent works \cite{Snoek,Snoek1}, quantum and thermal
fluctuations in finite vortex arrays in a one-dimensional optical lattice
were considered. See also Refs. \cite{Pogo1,Pogo2} for thermal fluctuations
in spinor condensates.

In the present paper, we study the intershell "melting" of small vortex
clusters in quasi-$2D$ BEC at different number of vortices in the system. We
consider the situation, when a cluster consists of only two shells. First,
we find the ground state configurations of vortices and then calculate
deviations of vortices from their equilibrium positions in a harmonic
approximation. We show that, if the number of vortices in the inner and
outer shell are not commensurate, deviations of shells with respect to each
other can be very significant even at low temperatures, $T\ll T_{c}$, and
large number of particles in the system, and shells become decoupled with
respect to each other thus leading to the disordering of the vortex cluster.
Also studied is the role of thermal fluctuations on the small cluster,
consisting on $2$, $3$, or $4$ vortices, in a trap with a small \textit{%
quadrupole deformation}, which acts as a source of orientational pinning for
the cluster.

The paper is organized as follows. In Section II we present our model, which
allows one to find an energetically favorable vortex configuration in the $%
2D $ case and also to calculate semi-quantitatively deviations of vortex
positions due to the thermal fluctuations. In Section III we study the
intershell "melting" process in different two-shells vortex clusters and
obtain a order-disorder transition temperature. In Section IV we analyze the
effect of thermal fluctuations on vortices in the trap with a small
quadrupole deformation. We conclude in Sec. V.

\section{Model}

Consider quasi-two-dimensional condensate with $N$ particles confined by the
radial harmonic trapping potential 
\begin{equation}
U(r)=\frac{m\omega _{\perp }^{2}r^{2}}{2},  \label{poten}
\end{equation}%
where $\omega _{\perp }$ is a trapping frequency, $m$ is the mass of the
atom, and $r$ is the radial coordinate. The system is rotated with the
angular velocity $\omega $. In this paper, we restrict ourselves on the
range of temperatures much smaller than $T_{c}$. Therefore, we can neglect a
noncondensate contribution to the free energy of the system. Thus, the
energy functional reads 
\begin{eqnarray}
F &=&\hbar \omega _{\perp }N(T)\int rdr\int d\varphi \Big(\frac{1}{2}|\nabla
\psi |^{2}+\frac{r^{2}}{2}|\psi |^{2}+2\pi g_{N}|\psi |^{4}  \notag \\
&&-i\omega \psi ^{\ast }\;\frac{\partial \psi }{\partial \phi }\Big),
\label{energi}
\end{eqnarray}%
where the integration is performed over the area of the system, $\varphi $
is the polar angle, $N(T)$ is the number of condensed atoms, $\omega $ is
the rotation frequency, $g_{N}=N\sqrt{\frac{2}{\pi }}\frac{a}{a_{z}}$ is the
interaction parameter, $a$ and $a_{z}$ are the scattering length and
oscillator length ($a_{z}=\sqrt{\frac{\hbar }{m\omega _{z}}}$) in $z$%
-direction, which is kinematically frozen. Distances and rotation
frequencies are measured in units of the radial oscillator length and the
trapping frequency, respectively. The normalization condition for the order
parameter reads $\int rdr\int d\varphi \left\vert \psi \right\vert ^{2}=1$.
In this paper, we analyze the case of dilute BEC and take $g_{N}=5$, which
corresponds to $\omega _{z}/2\pi =1.05$ kHz at $N=1000$ for $^{87}$Rb ($%
a\simeq 5.3$ nm). Since we consider the range of low temperatures, $%
T\lesssim 0.1T_{c}$, we can assume that $N(T)\simeq N$. For the dependence
of $T_{c}$ on $N$, we use the ideal gas result for the 2D case:

\begin{equation}
\frac{\hbar \omega _{\perp }}{kT_{c}}=\sqrt{\frac{\zeta (2)}{N}},
\end{equation}%
where $\zeta (2)$\ is a Riemann zeta function, $\sqrt{\zeta (2)}\approx 1.28$%
. Eq. (3) remains accurate even for the case of interacting particles \cite%
{Gies}.

\subsubsection{Ground state}

Now we present a method allowing one to find a ground state of the system,
which corresponds to the certain vortex cluster, and deviations of vortices
from their equilibrium positions due to thermal fluctuations.

\textit{In general case}, $\psi $ can be represented as a Fourier expansion%
\begin{equation}
\psi (r,\varphi )=\sum_{l}f_{l}(r)\exp (-il\varphi ).  \label{Fur'e}
\end{equation}%
Let us denote a number of vortices in the system as $v$. If the superfluid
phase in BEC has a $q$-fold symmetry, than only terms with $l$'s divisible
by $q$ survive in expansion (4). For instance, a vortex cluster consisting
of a single ring of $v$ vortices corresponds to the expansion (4) with $l=0$%
, $v$, $2v$, $3v$, .... A two-shell cluster with $v_{1}$ and $v_{2}$
vortices in the shells ($v_{1}+v_{2}=v$), where $v_{2}$\ is divisible by $%
v_{1}$, corresponds to the expansion (4) with $l=0$, $v_{1}$, $2v_{1}$, $%
3v_{1}$, ... If $v_{2}$\ is not divisible by $v_{1}$, than, in general case,
expansion (4) contains all harmonics. Typically, main contribution to the
energy is given by just few harmonics and by taking into account
approximately ten of them, one can find the energy of the system with very
high accuracy provided that the number of vortices in the cloud is not too
large, $v\lesssim 10\div 20$.

In the limit of noninteracting gas ($g_{N}=0$), it follows from the
Gross-Pitaevskii equation that each function $f_{l}$ coincides with the
eigen function for the harmonic oscillator corresponding to the angular
momentum $l$. These functions have the Gaussian profile $\sim r^{l}\exp
\left( -\frac{r^{2}}{2}\right) $. Therefore, one can assume that this
Gaussian approximation remains accurate in the case of weakly interacting
dilute gas. The accuracy can be improved if we introduce a variational
parameter $R_{l}$ characterizing the spatial extent of $f_{l}$. Finally, our 
\textit{ansatz} for $f_{l}$ has a form: 
\begin{equation}
f_{l}(r,C_{l},R_{l},\phi _{l})=C_{l}\left( \frac{r}{R_{l}}\right) ^{l}\exp
\left( -\frac{r^{2}}{2R_{l}^{2}}-i\phi _{l}\right) ,
\end{equation}%
where $C_{l}$, $R_{l}$, and $\phi _{l}$ can be found from the condition of
minimum of the energy (2), $C_{l}$ is a real number. This approach was used
for the first time in Ref. \cite{Rokshar} to evaluate energies and density
plots of different vortex configurations. In Ref. \cite{Kavoulakis}, the
simplified version of this method with fixed values of $R_{l}=1$ was applied
for the limit of weakly interacting gas with taking into account up to $9$
terms in the expansion (4). In Ref. \cite{Kavoulakis1}, the results for such
approximate solutions to the Gross-Pitaevskii equation were compared with
some known results of numerical solutions. A good accuracy of the ansatz was
revealed. See also Ref. \cite{Lundh} with the related approach. In Ref. \cite%
{Pogosov}, a version of this method was also used to calculate the energy of
axially-symmetric vortex phases in spinor condensates with the comparison of
the obtained results with numerical solutions, and a good agreement was
found. Therefore, this method can be also applied for our problem and we
expect that the results must be semi-quantitatively accurate and with the
help of this model one can reveal the effect of symmetry of vortex cluster
on melting temperatures and estimate values of those temperatures.

Now we substitute Eqs. (4) and (5) to Eq. (2) and after the integration we
obtain:

\ 
\begin{eqnarray}
\frac{F}{\hbar \omega _{\perp }N} &=&\sum_{l}\alpha
_{l}C_{l}^{2}+\sum_{l}I_{llll}C_{l}^{4}+4\sum_{l>k}I_{llkk}C_{l}^{2}C_{k}^{2}
\notag \\
&&+4\sum_{l>k>m}I_{lkkm}C_{l}C_{k}^{2}C_{m}\delta _{l+m,2k}  \notag \\
&&\times \cos \left( \phi _{l}+\phi _{m}-2\phi _{k}\right) \smallskip  \notag
\\
&&+8\sum_{l>k>m>n}I_{lkmn}C_{l}C_{k}C_{m}C_{n}\delta _{l+k,m+n}  \notag \\
&&\times \cos \left( \phi _{l}+\phi _{k}-\phi _{m}-\phi _{n}\right) ,
\label{ener}
\end{eqnarray}%
where 
\begin{equation}
\alpha _{l}=\frac{\pi }{2}\Gamma (l+2)\left( 1+R_{l}^{4}\right) +\pi
R_{l}^{2}\Gamma (l+1)\omega l,  \label{alfa}
\end{equation}%
\begin{eqnarray}
I_{lkmn} &=&2\pi ^{2}g_{N}\Gamma (\frac{l+m+n+k}{2}+1)R_{lkmn}^{2}  \notag \\
&&\times \left( \frac{R_{lkmn}}{R_{l}}\right) ^{l}\left( \frac{R_{lkmn}}{%
R_{k}}\right) ^{k}  \notag \\
&&\times \left( \frac{R_{lkmn}}{R_{m}}\right) ^{m}\left( \frac{R_{lkmn}}{%
R_{n}}\right) ^{n},  \label{integ}
\end{eqnarray}%
\begin{equation}
R_{lkmn}=\sqrt{2}\left( R_{l}^{-2}+R_{k}^{-2}+R_{m}^{-2}+R_{n}^{-2}\right)
^{-\frac{1}{2}},  \label{radius}
\end{equation}%
$\Gamma (l)$ is a gamma function. Normalization condition is now given by 
\begin{equation}
\pi \sum_{l}C_{l}^{2}R_{l}^{2}\Gamma (l+1)=1.  \label{normalization}
\end{equation}%
Values of parameters $R_{l}$, $C_{l}$\ and $\phi _{l}$\ can be found from
the minimum of the energy (6) taking into account Eq. (10). For instance,
for the axially-symmetric vortex-free state, $C_{0}=\sqrt{1/\pi R_{0}^{2}}$, 
$R_{0}=\left( 1+2g\right) ^{1/4}$, and $C_{l}=0$ at $l\eqslantgtr 1$. Note
that the energy is proportional to $\hbar \omega _{\perp }N$ at given values
of $\omega $\ and $g_{N}$.

\subsubsection{Thermal fluctuations: harmonic approximation}

After finding of ground state values of variational parameters, one can
calculate the equilibrium positions of vortices $\left\{ r_{0}^{(j)},\varphi
_{0}^{(j)}\right\} $, $j=1,...v,$ by numerical solution of equation\ 

\begin{equation}
\psi (r_{0}^{(j)},\varphi _{0}^{(j)},p_{n}^{(0)})=0,  \label{12}
\end{equation}%
where we introduced a notation $\left\{ p_{n}\right\} $ for the set of all
variational parameters ($R_{l}$, $C_{l}$\ and $\phi _{l}$) and $\left\{
p_{n}^{(0)}\right\} $\ denotes ground state values of these parameters.
Fluctuations of $p_{n}$, which are the degrees of freedom for the system in
this model, lead to the fluctuations of vortex positions. We denote
deviations of variational parameters from their equilibrium values as $%
\delta p_{n}$ and express deviations of vortices $\delta r^{(j)}$ and $%
\delta \varphi ^{(j)}$ through deviations of variational parameters in a
linear approximation. Perturbed positions of vortices are determined by the
equation 
\begin{equation}
\psi (r_{0}^{(j)}+\delta r^{(j)},\varphi _{0}^{(j)}+\delta \varphi
^{(j)},p_{n}^{(0)}+\delta p_{n})=0.  \label{13}
\end{equation}

Finally, deviation of the position of a given vortex from the equilibrium is%
\begin{equation}
\delta r^{(j)}=\frac{A_{n}^{(j)}}{D^{(j)}}\delta p_{n},  \label{14}
\end{equation}%
\begin{equation}
\delta \varphi ^{(j)}=\frac{B_{n}^{(j)}}{D^{(j)}}\delta p_{n}.  \label{15}
\end{equation}

Here and below repeated indices are summed; $A_{n}^{(j)}$, \ $B_{n}^{(j)}$,
and $D^{(j)}$\ are given by

\begin{equation}
A_{n}^{(j)}=Im\left( \frac{\partial \psi }{\partial p_{n}}\right) Re\left( 
\frac{\partial \psi }{\partial r}\right) -Re\left( \frac{\partial \psi }{%
\partial p_{n}}\right) Im\left( \frac{\partial \psi }{\partial r}\right) ,
\label{16}
\end{equation}

\begin{equation}
B_{n}^{(j)}=Im\left( \frac{\partial \psi }{\partial p_{n}}\right) Re\left( 
\frac{\partial \psi }{\partial \varphi }\right) -Re\left( \frac{\partial
\psi }{\partial p_{n}}\right) Im\left( \frac{\partial \psi }{\partial
\varphi }\right) ,  \label{17}
\end{equation}

\begin{equation}
D^{(j)}=Im\left( \frac{\partial \psi }{\partial r}\right) Re\left( \frac{%
\partial \psi }{\partial \varphi }\right) -Re\left( \frac{\partial \psi }{%
\partial r}\right) Im\left( \frac{\partial \psi }{\partial \varphi }\right) .
\label{18}
\end{equation}

All the derivatives on $p_{n}$, $r$, and $\varphi $\ in Eqs. (15)-(17) are
taken at ground state values of parameters $p_{n}=p_{n}^{(0)}$ and space
coordinates, corresponding to the equilibrium position of a given vortex, $%
r=r_{0}^{(j)}$, $\varphi =\varphi _{0}^{(j)}$. The squared deviation of
radial and polar coordinates of two vortices labelled as $j_{1}$ and $j_{2}$
with respect to each other is given by

\begin{equation}
\delta r_{(j_{1}j_{2})}^{2}=G_{mn}^{(j_{1}j_{2})}\delta p_{m}\delta p_{n},
\label{19}
\end{equation}

\begin{equation}
\delta \varphi _{(j_{1}j_{2})}^{2}=J_{mn}^{(j_{1}j_{2})}\delta p_{m}\delta
p_{n},  \label{20}
\end{equation}

where 
\begin{equation}
G_{mn}^{(j_{1}j_{2})}=\left( \frac{A_{n}^{(j1)}}{D^{(j1)}}-\frac{A_{n}^{(j2)}%
}{D^{(j2)}}\right) \left( \frac{A_{m}^{(j1)}}{D^{(j1)}}-\frac{A_{m}^{(j2)}}{%
D^{(j2)}}\right) ,  \label{21}
\end{equation}%
\begin{equation}
J_{mn}^{(j_{1}j_{2})}=\left( \frac{B_{n}^{(j1)}}{D^{(j1)}}-\frac{B_{n}^{(j2)}%
}{D^{(j2)}}\right) \left( \frac{B_{m}^{(j1)}}{D^{(j1)}}-\frac{B_{m}^{(j2)}}{%
D^{(j2)}}\right) .  \label{22}
\end{equation}

In the same manner, we can express deviations of energy from the ground
state value as a quadratic function in terms of deviations of variational
parameters:%
\begin{equation}
\delta F=E_{st}\delta p_{s}\delta p_{t},  \label{23}
\end{equation}

where%
\begin{equation}
E_{st}=\frac{\partial ^{2}F}{\partial p_{s}\partial p_{t}}.  \label{24}
\end{equation}%
The derivatives here are also calculated at $p_{n}=p_{n}^{(0)}$. 

The averaged squared deviations of radial and polar coordinates of two
vortices with respect to each other due to thermal fluctuations are given by

\begin{equation}
\left\langle \delta r_{(j_{1}j_{2})}^{2}\right\rangle _{T}=\frac{\int
d(\delta p)G_{mn}^{(j_{1}j_{2})}\delta p_{m}\delta p_{n}\exp (-\frac{1}{kT}%
E_{st}\delta p_{s}\delta p_{t})}{\int d(\delta p)\exp (-\frac{1}{kT}%
E_{st}\delta p_{s}\delta p_{t})},  \label{25}
\end{equation}

\begin{equation}
\left\langle \delta \varphi _{(j_{1}j_{2})}^{2}\right\rangle _{T}=\frac{\int
d(\delta p)J_{mn}^{(j_{1}j_{2})}\delta p_{m}\delta p_{n}\exp (-\frac{1}{kT}%
E_{st}\delta p_{s}\delta p_{t})}{\int d(\delta p)\exp (-\frac{1}{kT}%
E_{st}\delta p_{s}\delta p_{t})}.  \label{26}
\end{equation}%
In general case, integrals in Eqs. (24) and (25) can not be calculated
analytically, since matrix $E_{st}$ is not necessarily diagonal. Therefore,
we have to switch to a new basis $\delta t=M\delta t$, where $M$ is a
matrix, which diagonalizes quadratic form (22). Here and below we will use a
matrix form for the equations. Quadratic forms (18), (19), (22) in the new
basis can be written as%
\begin{equation}
\delta r_{(j_{1}j_{2})}^{2}=\left( \delta t\right)
^{T}P^{(j_{1}j_{2})}\delta t,  \label{27}
\end{equation}%
\begin{equation}
\delta \varphi _{(j_{1}j_{2})}^{2}=\left( \delta t\right)
^{T}R^{(j_{1}j_{2})}\delta t,  \label{28}
\end{equation}%
\begin{equation}
\delta F=\left( \delta t\right) ^{T}Q\delta t,  \label{29}
\end{equation}%
where $Q=M^{T}EM$, $P^{(j_{1}j_{2})}=M^{T}G^{(j_{1}j_{2})}M$, and $%
R^{(j_{1}j_{2})}=M^{T}J^{(j_{1}j_{2})}M$. Matrix $Q$ must be diagonal and
from this condition one can find matrix $M$ numerically and then calculate $%
P^{(j_{1}j_{2})}$\ and $R^{(j_{1}j_{2})}$. In the new basis, integrals in
Eqs. (24) and (25) can be found analytically and finally we get: 
\begin{equation}
\left\langle \delta r_{(j_{1}j_{2})}^{2}\right\rangle _{T}=kT\frac{%
P_{nn}^{(j_{1}j_{2})}}{Q_{nn}},  \label{30}
\end{equation}%
\begin{equation}
\left\langle \delta \varphi _{(j_{1}j_{2})}^{2}\right\rangle _{T}=kT\frac{%
R_{nn}^{(j_{1}j_{2})}}{Q_{nn}}.  \label{31}
\end{equation}%
As usual in a harmonic approximation, average squares of deviations are
proportional to the temperature.

For the vortex cluster consisting of two shells we also introduce a quantity 
$\Delta \varphi $, which has a sense of averaged displacement of vortex
shells with respect to each other. It can be defined as a root of square
displacement of pair of vortices from different shells averaged over all
possible pairs of vortices: 
\begin{equation}
\Delta \varphi =\left[ \frac{1}{v_{1}v_{2}}\sum_{j_{1},j_{2}}\left\langle
\delta \varphi _{(j_{1}j_{2})}^{2}\right\rangle _{T}\right] ^{1/2}.
\label{32}
\end{equation}%
Now, if we take into account Eq. (31) and the fact that energy in the ground
state is proportional to $\hbar \omega _{\perp }N$, we obtain the following
relation:

\begin{equation}
\Delta \varphi =\frac{t^{1/2}}{N^{1/4}}d(g_{N},\omega ).  \label{33}
\end{equation}%
where $t$ is a reduced temperature, $t=T/T_{c}$; function $d(g_{N},\omega )$%
\ depends on interaction constant $g_{N}$\ and rotation speed $\omega $. Of
course, $d(g_{N},\omega )$ is also very strongly dependent on the vortex
cluster symmetry and in the next Section we will calculate it for some
values of $g_{N}$\ and $\omega $ and vortex configurations.

Note that the harmonic approximation remains accurate only if the deviations
of the positions of vortices are much smaller than the characteristic
distance between two neighboring vortices. The "melting" temperature can be
defined through the Lindemann criterion.

\section{Intershell melting of vortex clusters}

If there are not many vortices in the system, they are situated in
concentric shells. In the single vortex state, a vortex occupies the center
of the cloud. If the number of vortices $v$ is more than one, but less than
six, vortices are arranged in one shell. With further increasing of number
of vortices, one of the vortices jumps to the center of the cloud, whereas
others are still situated in the single shell \cite{Fetter}. Thermal
fluctuations in these cases can lead only to the radial displacements of
vortex positions, since there is only one shell in the system. However, when
the number of vortices is increased, they are arranged in two shells. For
instance, it was shown in Ref. \cite{Rokshar}, that in the phase with ten
vortices, two of them are situated in the inner shell and eight are arranged
in the outer shell. For phases with larger amount of vortices, their number
in the inner shell can increase.

Here, we consider the process of the intershell disordering in two-shells
clusters containing $10$, $11$, $12$, and $13$ vortices, respectively. It
would be more convenient for the comparison to calculate "melting"
temperatures for these configurations at the same value of rotation
frequency. However, only one of these states can be a true ground state and
if the system is not in a ground state, than sooner or later it will switch
to the ground state due to thermal fluctuations. Therefore, we find
"melting" temperatures for different vortex configurations at different, but
quite close to each other rotation speeds, which correspond to ground states
of the given configuration.

We choose the value of the gas parameter $g_{N}=5$, as was explained in
Section II, and find the ground states of the system. We have obtained that
a two-shells vortex clusters consisting of $v=10$, $11$, $12$, and $13$
vortices are energetically favorable in the vicinity of the point $\omega
=0.9$. For instance, the ground state of the system is represented by phases
with $10$, $11$, $12$, and $13$ vortices at $\omega =0.9$, $0.91$, $0.92$, $%
0.94$, respectively. In these cases, the inner shells contain $v_{1}=2$, $3$%
, $3$, and $4$ vortices, whereas the outer shells have $v_{2}=8$, $8$, $9$,
and $9$ vortices, respectively. Density plots for these vortex phases are
shown in Fig. 1. Let's calculate deviations of vortex positions for these
states.

Using a technique presented in the previous Section, we found that if we
increase the temperature from zero, at first, fluctuations of relative
phases $\phi _{l}$ of different harmonics of the order parameter becomes
important and deviations of the positions of vortices are almost entirely
due to the fluctuations of $\phi _{l}$ and not due to fluctuations of $C_{l}$
and $R_{l}$. This can be expected, since it is well-known that fluctuations
of the phase of the order parameter are more pronounced at relatively low
temperatures and only at much higher temperatures an amplitude of the order
parameter starts to fluctuate. Also, fluctuations of $\phi _{l}$\ lead
predominantly to azimuthal displacements of vortices, displacements in a
radial direction are much smaller. This reflects the fact that the
temperature of the intershell melting is much lower than that of the radial
melting. We define an intershell melting temperature $t_{melt}$ of the
cluster as a temperature, at which $\Delta \varphi $ is equal to $\gamma 
\frac{360}{v_{2}}$, where $\gamma \sim 0.1$ is a characteristic number from
the Lindemann criterion:

\begin{equation}
t_{melt}=\left( \frac{360}{n_{2}d(g_{N},\omega )}\gamma \right) ^{2}\sqrt{N}.
\label{34}
\end{equation}%
Factor $360/v_{2}$ in Eq. (33) reflects the fact that the two-shell cluster
is invariant under the rotation of shells with respect to each other on
angle $360%
{{}^\circ}%
/v_{2}$. We have calculated values of $d$ for clusters with $10$, $11$, $12$%
, and $13$ vortices. Our results are $d(g_{N},\omega )\approx 608%
{{}^\circ}%
$, $3500%
{{}^\circ}%
$, $123%
{{}^\circ}%
$, and $810%
{{}^\circ}%
$ for $10$, $11$, $12$, and $13$ vortex clusters, respectively. We can see
that $10$ and $12$ vortex clusters are most stable among the analyzed
configurations, and average angle between the shells is less than in other
cases. This is because the number of vortices in the outer shell $v_{2}$ is
divisible by $v_{1}$. Intuitively, it is clear that the stability of a
cluster with $v_{2}$ divisible by $v_{1}$ depends also on the ratio $%
v_{2}/v_{1}$, since in the limit $v_{2}/v_{1}\rightarrow 1$, each vortex in
the inner shell corresponds to one vortex from the outer shell. Probably,
this is a reason, why $12$-vortex cluster is more stable than $10$-vortex
configuration. At the same time, $11$ and $13$-vortex clusters are the most
unstable among considered here, since $v_{2}$ and $v_{1}$ are
incommensurate, and a deviation of shells with respect to each other is the
largest. Note that with changing of $\omega $ with fixed $g_{N}$, $%
d(g_{N},\omega )$ increases, in accordance with calculations \cite{Tkachenko}
for Tkachenko modes. One can see from Eq. (34) and our estimates for $%
d(g_{N},\omega )$ that $12$-vortex cluster is not melted and remains stable
at $N=10^{3}$ and $t\approx 0.1$, whereas in other cases a displacement
angle between different shells is comparable with the angle between the two
neighboring vortices in the outer shell and therefore the shells are
decoupled. The difference in melting temperatures for 12- and 11-vortex
clusters is several orders of magnitude.

Experimentally, melting of vortex clusters can be studied by tuning of $%
\omega $ at fixed $g_{N}$ and $N$. After obtaining a desirable vortex
configuration, one can also tune $T$\ and reach a "melting" range of
temperatures. Vortex positions can be found by the free expansion technique
and after the repeating of this procedure one can determine the average
deviation of the vortex positions from the equilibrium. It is also possible
to use a Bragg spectroscopy for the systems containing much larger vortex
arrays than those considered here.

\section{Vortices in a trap with quadrupole deformation}

In this Section we consider the effect of thermal fluctuations on vortices
in the trap with a quadrupole deformation, which breaks a rotational
symmetry. Such a deformation is often used in experiments to facilitate the
creation of vortices. In fact, it introduces a preferable direction for the
arrangements of vortices acting as a source of orientational pinning for a
vortex cluster. The additional quadrupolar pinning potential is given by%
\begin{equation}
U_{quadr}(r)=\frac{\varepsilon m\omega _{\perp }^{2}r^{2}\cos 2\varphi }{2},
\label{35}
\end{equation}%
where $\varepsilon $ is a small coefficient, $\varepsilon \ll 1$. At zero
temperature quadrupole deformation of the trap potential fixes azimuthal
positions of vortices, whereas thermal fluctuations lead to displacements of
vortices, which depend on $\varepsilon $ and $T$. Note that, in the case of
a two-shell cluster at $\varepsilon =0$, considered in the previous Section,
pinning centers are created by each vortex shell for another shell, and the
total rotational symmetry is preserved. Additional potential (34) leads to
the following contribution to the energy (6):

\begin{eqnarray}
\frac{F_{quadr}}{\hbar \omega _{\perp }N} &=&2\pi \varepsilon
\sum_{m}c_{m}c_{m+2}\frac{R_{m+2}^{m+4}R_{m}^{m+6}}{\left(
R_{m+2}^{2}+R_{m}^{2}\right) ^{m+3}}  \notag \\
&&\times \Gamma (m+3)\cos (\phi _{m+2}-\phi _{m}).
\end{eqnarray}%
This term relates phases and amplitudes of different harmonics of the order
parameter with angular momenta different by $2$ from each other. Here we
analyze the situation, when there are $v=2$, $3$ or $4$ vortices in the
trap. Density plots for these vortex states are presented in Fig. 2. The
direction of minimum of trapping potential is vertical. We found that at
small quadrupole deformation, $\varepsilon \lesssim 0.1$, fluctuations of
relative phases of the order parameter Furrier harmonics are much stronger
than fluctuations of their amplitudes, similarly to the case considered in
the previous Section. Therefore, we apply the same ideas to the present
problem. Fluctuations of the relative phases lead mostly to the
displacements of vortices in the azimuthal direction. These are so called
scissors modes, which are responsible for such oscillations \cite%
{Stringari,Foot,Zaremba,Castin}.

In the limit of small $\varepsilon $, one can consider $F_{quadr}$ as a
perturbation to the energy of the system with $\varepsilon =0$. If there are
two vortices in the system, the nonperturbed order parameter contains all
harmonics divisible by $2$, and the main contribution to the energy is given
by harmonics with $l=0$ and $2$. Amplitudes of these harmonics are of $%
\varepsilon ^{0}$ order of magnitude, relative angles between them are
fixed, and the energy is degenerate with respect to $\phi _{2}$, which
reflects the fact that vortices can rotate freely (Goldstone mode).
Quadrupole deformation connects $\phi _{0}$ and $\phi _{2}$, $\phi _{2}$ and 
$\phi _{4}$, $\phi _{4}$ and $\phi _{6}$, etc.. As a result, $F_{quadr}\sim $
$\varepsilon ^{1}$ and the angle of deviation of vortex cluster due to
thermal fluctuations is given by 
\begin{equation}
\Delta \varphi =\frac{t^{1/2}}{\varepsilon ^{\tau }N^{1/4}}d(g_{N},\omega ),
\label{37}
\end{equation}%
where $d(g_{N},\omega )$\ is a function, independent on $t$; $\tau =0.5$.

If there are $4$ vortices in the system, than at $\varepsilon =0$ the order
parameter contains all the harmonics divisible by $4$. However, these
harmonics cannot be related through the Eq. (35), since their angular
momenta should differ by $2$ and not by $4$. In this case, coupling of
vortex cluster to the quadrupole deformations occurs in the next order of $%
\varepsilon $. Namely, quadrupole deformation induces a harmonic with $l=2$,
whose amplitude is of the order of $\varepsilon ^{1}$, and finally $%
F_{quadr}\sim $ $\varepsilon ^{2}$; and we again arrive to Eq. (36) with $%
\tau =1$.

Now we consider the situation, when there are $3$ vortices in the system.
Again, a nonperturbed order parameter consists of contributions with $l$'s
divisible by $3$, and their phases are not related by Eq. (35), as in the
previous case with $4$ vortices. In the next approximation with respect to $%
\varepsilon $, quadrupole deformation induces other harmonics with all
integer $l$'s, and amplitudes of these harmonics are of the order of $%
\varepsilon ^{1}$. This is possible, because any integer $l$, which is not
divisible by $3$, can be represented as $3l\pm 2$, and therefore it can be
obtained by adding or substraction $2$ from $3l$. However, it turns out that
the energy even in the second order of $\varepsilon $ is again degenerate
and this degeneracy is removed only in the next order of $\varepsilon $.
After all, $F_{quadr}\sim $ $\varepsilon ^{3}$ and Eqs. (36) is again valid
but with $\tau =1.5$.

We see that the symmetry of the vortex configuration dictates the asymptotic
behavior of the pinning energy and the average deviation of vortex cluster
at $\varepsilon \rightarrow 0$. The cluster with $2$ vortices is most
strongly pinned, whereas the cluster with $3$ vortices is the most unstable.
This effect reflects the commensurability of the angular momenta of the
quadrupole deformation ($l=2$) and of the order parameter harmonics,
responsible for vortices ($l=2$, $3$, $4$). The strongest pinning is
observed, when these momenta are equal to each other (two-vortex state),
less stronger pinning, when these momenta are commensurate, but not equal
(four-vortex phase), and the most weak pinning for incommensurate case
(three-vortex state). Note that similar scaling relations for the frequences
of scissors modes in $2$ and $3$ vortex states were obtained recently in
Ref. \cite{Castin}.

Next we calculate values of $d(g_{N},\omega )$ for $2$, $3$ and $4$ vortex
cluster at $g_{N}=5$ and $\omega =0.68$, $0.75$, and $0.78$, respectively,
where these vortex configurations are energetically favorable, according to
our calculations. Our results are $d(g_{N},\omega )\approx 65%
{{}^\circ}%
$, $130%
{{}^\circ}%
$, and $11%
{{}^\circ}%
$ for $2$, $3$, and $4$ vortex states, respectively. If the deviation angle $%
\Delta \varphi $ becomes of the same order as the angle between two
neighboring vortices in the cluster, $2\pi /v$, we will treat this vortex
cluster as being depinned from the quadrupole deformation. By using Eq. (36)
and this condition, one can easily obtain phase diagram in the $(t,$ $%
\varepsilon )$ space. The example of the phase diagram is presented on Fig.
3 for $g_{N}=5$ and $N=1000$. Each line determines the boundary between the
pinned and unpinned vortex cluster for a given vortex configuration. Below
these lines, vortex cluster is pinned and above it is unpinned. One can see
that the region of stability of $2$-vortex cluster is much broader than that
for two other configurations.

\section{Conclusions}

In this paper, we studied the effect of thermal fluctuations on small vortex
clusters in harmonically trapped rotating Bose-Einstein condensate at
temperatures much lower than the critical temperature. First, we have
considered the clusters, consisting of two concentric shells of vortices.
These were $10$, $11$, $12$, and $13$ vortex structures. We obtained that
with increasing the temperature from zero, first, an order between the
positions of vortices from different shells is destroyed, whereas the order
within each shell is preserved. By using a Lindemann criterion, we defined
the temperature, corresponding to the decoupling of two shells of vortices
with respect to each other, which determines an order-disorder transition.
This "melting" temperature is strongly dependent on the commensurability of
the number of vortices in shells; less commensurate clusters have lower
"melting" temperature. For instance, the "melting" temperatures for the $11$%
-vortex cluster consisting of two shells with $3$ and $8$ vortices and for
the commensurate $12$-vortex cluster with $3$ and $9$ vortices in shells
differ in several orders of magnitude. Intershell order-disorder transition
can be observed at experimentally attainable range of parameters. \textit{We
have shown that intershell melting in atomic condensates can occur at very
low temperatures, especially for incommensurate clusters.}

Also studied are vortex clusters in the trap with small quadrupole
deformation of the trapping potential, which acts as an orientational
pinning center for vortices. We have analyzed the case of $2$, $3$ and $4$
vortices in the system. We have demonstrate that the pinning energy depends
very strongly on the number of vortices in the system. With tending the
quadrupole deformation $\varepsilon $ to zero, the pinning energy becomes
proportional to $\varepsilon ^{\upsilon }$, where coefficient $\upsilon =1$, 
$3$, and $2$ for $2$, $3$, and $4$ vortex configurations, respectively. This
is due to the commensurability between the angular momenta $l=2$ transferred
to the system by the quadrupole deformation and $l=2$, $3$, and $4$,
responsible for the creation of $2$, $3$, and $4$ vortices, respectively.
Average deviation angles between the vortex cluster and the trap anisotropy
direction diverge in different power laws with tending $\varepsilon $ to
zero for different vortex configurations, $\Delta \varphi \sim \varepsilon
^{\tau }$, with $\tau =0.5$, $1.5$, and $1$ for $2$, $3$, and $4$ vortex
clusters, respectively.

\acknowledgments

Authors acknowledge useful discussions with T. K. Ghosh. W. V. Pogosov is
supported by the Japan Society for the Promotion of Science.

\section{Figure captions}

\textbf{Fig. 1 }(Color online). Density plots for the states with $10$, $11$%
, $12$, and $13$ vortices. Dark spots correspond to vortices.

\textbf{Fig. 2 }(Color online). Density plots for the states with $2$, $3$,
and $4$ vortices in a trap with small quadrupole deformation. Dark spots
correspond to vortices.

\textbf{Fig. 3}. The phase diagram of the $2$, $3$, and $4$-vortex clusters
in the trap with the quadrupole deformation at $N=1000$. Above these lines
the cluster is decoupled from the deformation and below it is coupled.

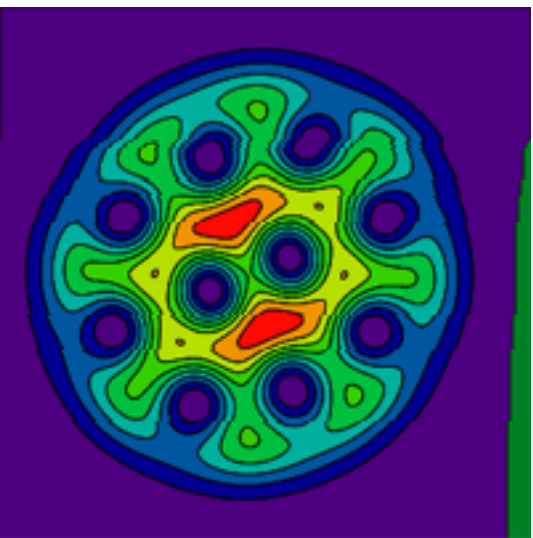
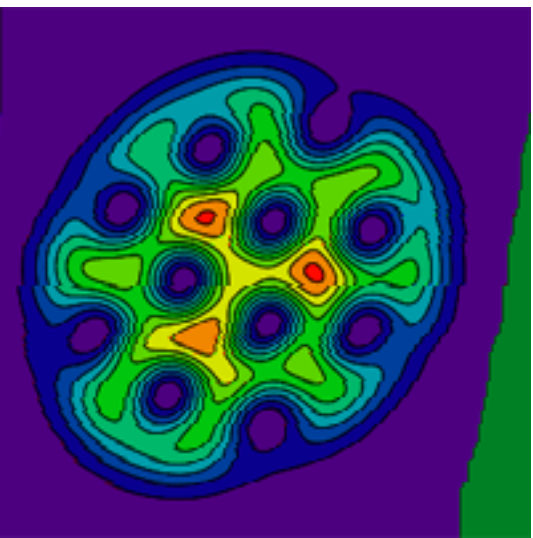
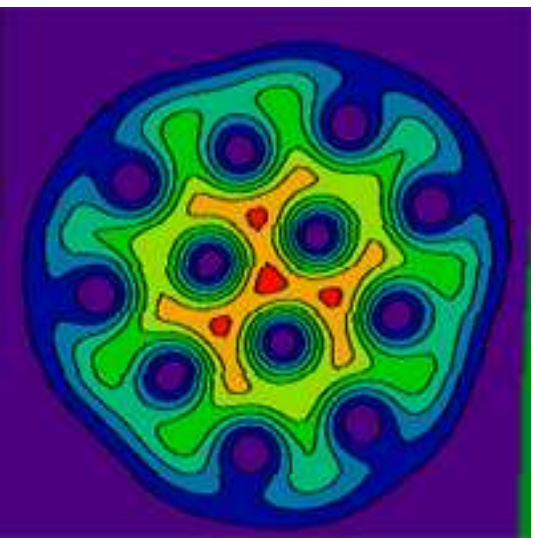
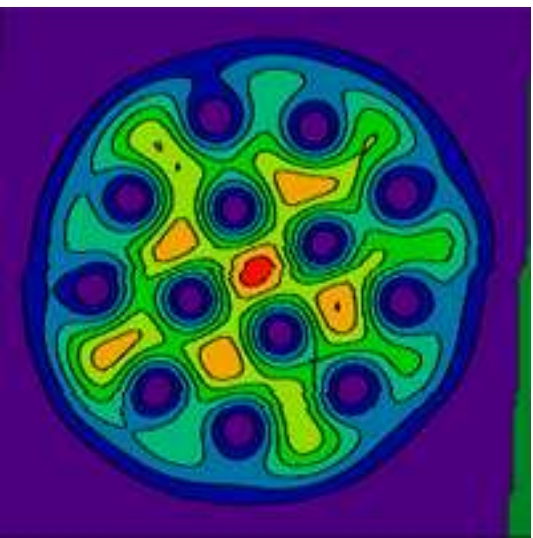
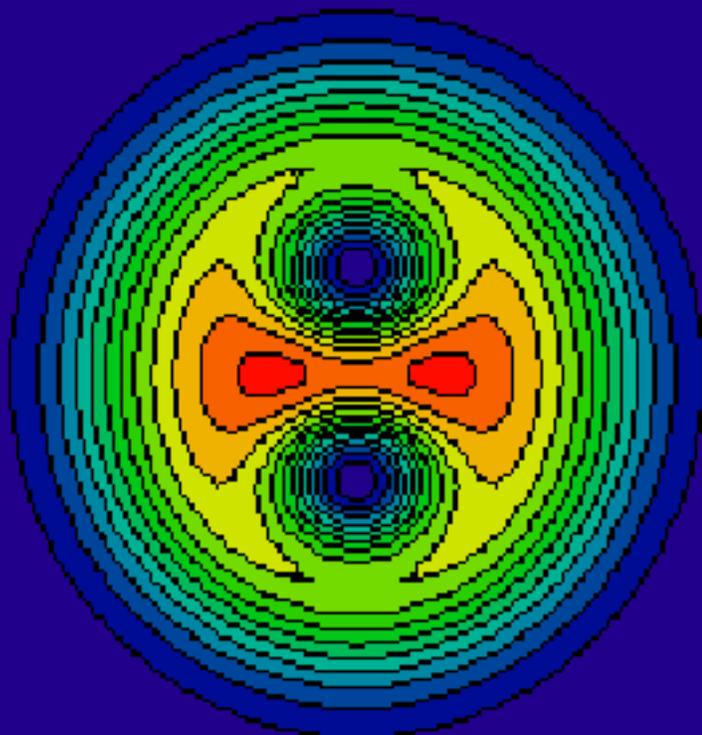
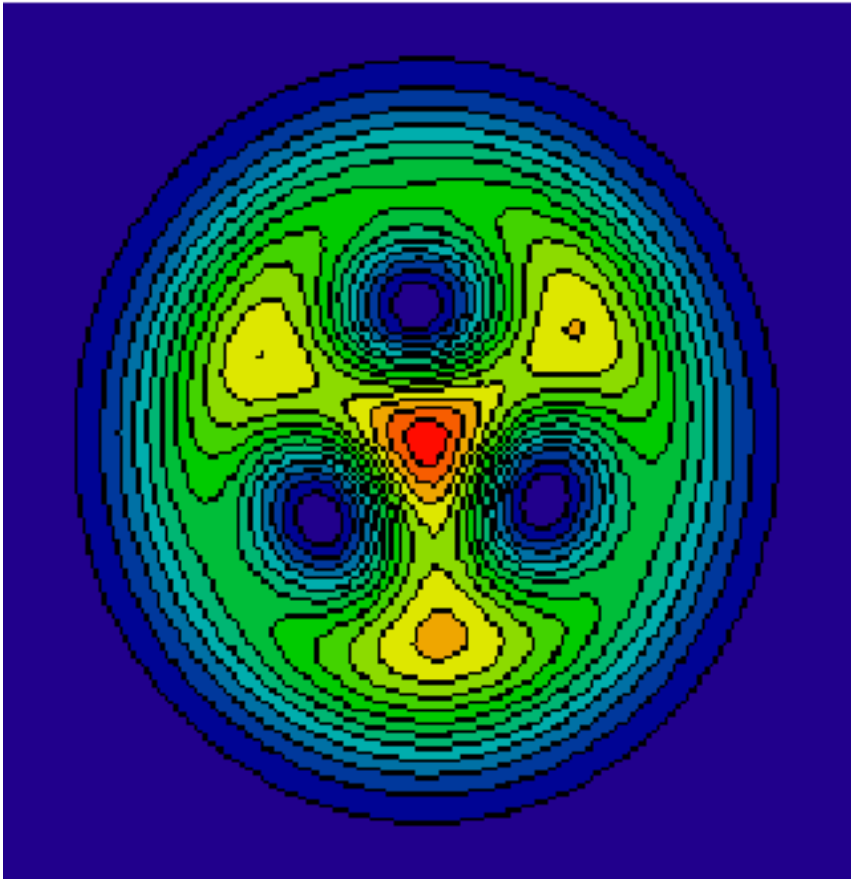
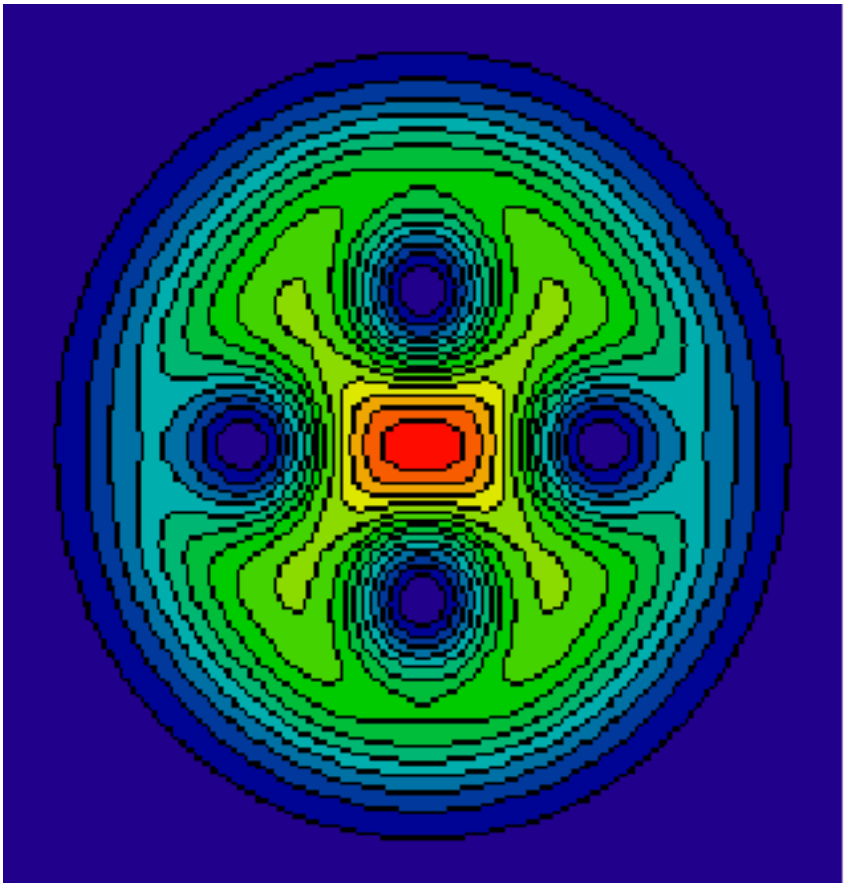
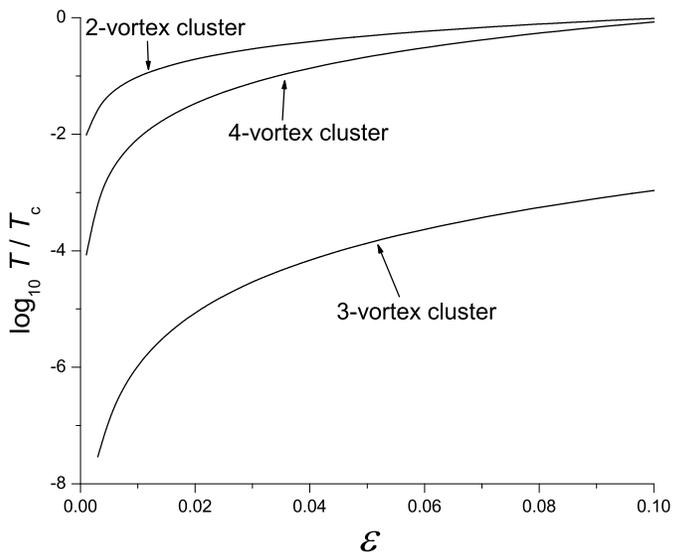
\end{document}